\documentclass{article}

\usepackage{arxiv}

\usepackage[utf8]{inputenc} 
\usepackage[T1]{fontenc}    

\RequirePackage{color}
\definecolor{todo}{rgb}{1,0,0}
\definecolor{answer}{rgb}{0,0,1}
\definecolor{new}{rgb}{1,0,1}
\definecolor{conditional}{rgb}{0,1,0}
\definecolor{e-mail}{rgb}{0,.40,.80}
\definecolor{reference}{rgb}{.20,.60,.22}
\definecolor{mrnumber}{rgb}{.80,.40,0}
\definecolor{citation}{rgb}{0,.40,.80}

\usepackage[colorlinks, linkcolor=reference,
 citecolor=citation,
         urlcolor=e-mail]{hyperref}
\usepackage{url}            
\usepackage{booktabs}       
\usepackage{bm,amsmath,amssymb,amsfonts}
\usepackage{mathtools}
\usepackage[usenames,dvipsnames]{xcolor}
\usepackage{graphicx}
\usepackage{caption}
\usepackage{subcaption}
\usepackage{listings}
\lstdefinelanguage{Julia}%
  {morekeywords={abstract,
  break, case, catch, const, continue, do, else, elseif, end, export, false, for, function, immutable, import, importall, if, in, macro, module, otherwise, quote, return, switch, true, try, type, typealias, using, while},%
   sensitive=true,%
   alsoother={$},%
   morecomment=[l]\#,%
   morecomment=[n]{\#=}{=\#},%
   morestring=[s]{"}{"},%
   morestring=[m]{'}{'},%
}[keywords,comments,strings]%

\DeclarePairedDelimiter\abs{\lvert}{\rvert}%
\DeclarePairedDelimiter\norm{\lVert}{\rVert}%
\makeatletter
\let\oldabs\abs
\def\abs{\@ifstar{\oldabs}{\oldabs*}}
\let\oldnorm\norm
\def\norm{\@ifstar{\oldnorm}{\oldnorm*}}
\makeatother
\usepackage{cleveref}
\usepackage{nicefrac}       
\usepackage{microtype}      
\usepackage{cleveref}       
\usepackage{lipsum}         
\usepackage{graphicx}
\usepackage{float}
\usepackage[square,numbers]{natbib}
\usepackage{makecell}

\usepackage{doi}
\usepackage{multicol,multirow}
\usepackage{tikz}
\usetikzlibrary{positioning}
\usetikzlibrary{calc}
\usetikzlibrary{shapes,arrows}

\title{Robust Parameter Estimation for \\Rational Ordinary Differential Equations}

\author{ 
{Oren Bassik} \\
Ph.D. Program in Mathematics\\
CUNY Graduate Center \\
	New York, NY, USA \\
	\texttt{orebas@yahoo.com} \\
	\And
 {Yosef Berman} \\
Ph.D. Program in Mathematics\\
CUNY Graduate Center \\
	New York, NY, USA \\
	\texttt{yberman1@gradcenter.cuny.edu} \\
	\And
{Soo Go} \\
Ph.D. Program in Computer Science\\
CUNY Graduate Center \\
	New York, NY, USA \\
	\texttt{sgo@gradcenter.cuny.edu} \\
	\And
    {Hoon Hong} \\
	Department of Mathematics\\
	North Carolina State University\\
	Raleigh, NC, USA\\
	\texttt{hong@ncsu.edu} \\
    \And
    {Ilia Ilmer} \\
    Ph.D. Program in Computer Science\\
    CUNY Graduate Center\\
    New York, NY, USA\\
    \texttt{iilmer@gradcenter.cuny.edu} \\
	\And
	Alexey Ovchinnikov \\
	Department of Mathematics\\
    CUNY Queens College\\
    Ph.D. Programs in Mathematics and Computer Science\\
    CUNY Graduate Center\\
    New York, NY, USA\\
    \texttt{aovchinnikov@qc.cuny.edu}\\
    \And
 {Chris Rackauckas} \\
Department of Mathematics\\Massachusetts Institute of Technology\\
	Cambridge, MA, USA \\
\texttt{crackauc@mit.edu}
	\And
    Pedro Soto\\
    Mathematical Institute\\ University of Oxford\\
    Oxford, UK\\
    \texttt{pedro.soto@maths.ox.ac.uk}\\
    \And
    Chee Yap\\
	Department of Computer Science,\\
    Courant Institute of Mathematical Science,\\
    New York University,\\
    New York, NY, USA\\
    \texttt{yap@cs.nyu.edu}\\
}

\hypersetup{
pdftitle={A template for the arxiv style},
pdfsubject={q-bio.NC, q-bio.QM},
pdfauthor={David S.~Hippocampus, Elias D.~Striatum},
pdfkeywords={},
}

\newcommand{\pn}[1]{{\footnotesize\texttt{#1}}}

\date{}

\begin{document}

\maketitle

\begin{abstract}
  We present a new approach for estimating parameters in  rational ODE models from given (measured) time series data.  
  
  In typical existing approaches, an initial guess for the parameter values is made from a given search interval. Then, in a loop, the corresponding outputs are computed by solving the ODE numerically, followed by computing the error from the given time series data. If the error is small, the loop terminates and the parameter values are returned. Otherwise, heuristics/theories are used to possibly improve the guess and continue the loop.
 
These approaches tend to be non-robust in the sense that their accuracy depend on the search interval and the true parameter values; furthermore, they cannot 
 handle the case where the parameters are locally identifiable.

  In this paper, we propose a new approach, which does not suffer from the above non-robustness.
  In particular, it does not require making good initial guesses for the parameter values or specifying search intervals. 
  Instead, it uses differential algebra, interpolation of the data using rational functions, and multivariate polynomial system solving.
  We  also compare the performance of the resulting software  with several other estimation software packages.
\end{abstract}

\keywords{Parametric ODE Models \and Parameter Estimation \and Differential Algebra \and Symbolic-Numeric Differentiation \and Mathematical Software}

\section{Introduction}
\subsection{Overall problem}
Parametric ODEs are ubiquitous in science and engineering.
In order to use them, one first needs to estimate the parameters, which is usually done using
measured time series data. 
Hence, the overall problem is: given a parametric ODE and
some measured data, to estimate the parameters.

\subsection{State of the art}
Due to its importance, this problem has been the subject of intensive research efforts  that have yielded various theories and approaches. They can be roughly categorized into three approaches: {\em shooting}, {\em two-stage},  and {\em algebraic}. 

{\em Shooting\/} approaches (also known as simulation-based approaches) directly use the ODE solver to compute the loss function of the current estimate. This is used to define a nonlinear optimization to improve the parameter estimates. This popular approach has seen many software implementations;
to list a few: AMIGO2, COPASI, Data2Dynamics, SBtoolbox2/IQM;
see  \cite{Bardbook,Beckbook,Ljungbook,WPBook,S2003,SBtoolbox,parameter_estimation,COPASI,FJS2008,EHCVMDBS2014,AMIGO2,Data2Dynamics,GVB2017,VFWHB2019,distefano2015dynamic}. Many studies have further developed the shooting approaches to improve robustness, adding techniques such as multiple shooting \cite{
bock2015direct,aydogmus2021modified,turan2021multiple}, automatic differentiation and adjoint techniques for more accurate derivatives \cite{9622796,navon1998practical,lakomiec2014parameter}, and landscape smoothing techniques like the prediction-error method \cite{larsson2009direct}, all of which help the optimization techniques avoid local minima.

{\em Two-stage} approaches (also known as collocation, smoothing, or principle differential analysis) avoid numerically solving the ODE by fitting collocation polynomials to the data to obtain predictions of the data which are then directly fit against the data \cite{ramsay1991some,campbell_2013}. As the result, these approaches are generally much more efficient. Previous studies have used various objects  such as B-splines \cite{10.1214/07-EJS132,poyton2006parameter,li2002estimation} and local least-squares \cite{liang2008parameter,roesch2021collocation,ramsay2007parameter, campbell2013maximum}. One downside to two-stage approaches is requiring observation of all states.  
Furthermore, it has been noted that the error introduced through the derivative estimation tends to lead to less accurate final results in real-world scenarios with less dense data \cite{poyton2006parameter,campbell_2013}.  These studies suggest finishing a parameter estimation by using the output of a two-stage approach as the initial condition to a simulation-based approach. 

{\em Algebraic} approaches tackle rational differential models by exploiting theoretical results from differential algebra. 
The paper~\cite{FJS2008} expresses parameters algebraically in terms of derivatives of input and output functions and then estimates these derivatives using the numerical data. Two difficulties with this approach are the size of these algebraic expressions and the computational complexity in obtaining them.
The papers
\cite{BKLPPU2014,JV2022,VMZD2020,ZVDK2018} study how to use input-output equations in parameter estimation and how to lower the orders of derivatives used in the estimation (because higher order derivatives typically introduce larger errors with numerical data) by different approaches, including using integrals, cf.~\cite[Section~3.2]{SGCL2014}. Computing input-output equations could be computationally very costly, though recently there have been improvements in algorithms computing them~\cite{DGHP2023}.
The paper~\cite{Q2017} considers an algebraic approach to parameter estimation in linear systems.
Finally, the paper ~\cite{KY2020} investigates the parameter manifold in the case where the differential model is structurally non-identifiable.

\subsection{The proposed approach and novelty}
The proposed approach could roughly be classified as a two stage approach. However, it is substantially different from the previous two stage approaches, in that it estimates parameters {\em robustly\/}, in the following sense: 
\begin{enumerate}\itemsep 0.5em
    \item The proposed approach produces good estimates no matter what the specified search intervals are.  In fact, it does not even require specifying any search interval.
    
    \item The proposed approach produces good estimates no matter where the true parameter values are.
    
    \item  The proposed approach produces good estimates even when the parameters are not globally identifiable, as long as they are locally identifiable.
\end{enumerate}
It achieves the robustness by judiciously combining:
\begin{itemize}\itemsep 0.5em
  \item differential algebra approach from the parameter identifiability analysis developed in~\cite{hong_sian_2019,hong_global_2020},
  \item estimation of the derivatives of the ODE solutions from rational interpolation of the data,
  \item overdetermined polynomial system solving by squaring the system and filtering the solution set.
\end{itemize}

\noindent As such, this technique can be used in isolation for dense data, or together with shooting approaches by providing more robust initial conditions for sparse data.

\subsection{Software Implementation}
We implemented the proposed approach into a Julia package called ``ParameterEstimation.jl''. All code and benchmarking data are available at {\footnotesize \url{https://github.com/iliailmer/ParameterEstimation.jl}}.

\section{Problem Statement}\label{subsec:est-approach}
In this section, we precisely state the  problem  solved by the proposed approach and illustrate it  using a toy example.
We also show how to enter/solve the problem using our soft implementation.

\medskip
\noindent\textbf{Input:}
\smallskip
\begin{enumerate} \itemsep 0.3em
  \item An ODE model $\Sigma$
        \begin{equation}\label{eq:main}
          \begin{cases}
            \bm{x}'(t) = \bm{f}(\bm{x}(t),  \bm{u}(t),\bm{\mu}) \\
            \bm{y}(t) = \bm{g}(\bm{x}(t),  \bm{u}(t),\bm{\mu})  \\
            \bm{x}(0) = \bm{x_0}
          \end{cases}
        \end{equation}
        where we use bold fonts for vector:
        \begin{itemize} \itemsep 0.3em
          \item $\bm{f}$ and $\bm{g}$ are 
                rational functions describing the model,
          \item $\bm{x}$ are state variables,
          \item $\bm{u}$ are input (control) variables,
          \item $\bm{y}$ are output variables,
          \item $\bm{\mu}$ and $\bm{x_0}$ are unknown  parameters.
        \end{itemize}
  \item A data $D=((t_1,\bm{y}_1),\ldots,(t_n,\bm{y}_n))$ where $\bm{y}_i$ is the measured value of~$\bm{y}$ at time~$t_i$
\end{enumerate}

\medskip

\noindent\textbf{Output:} Estimated values for the parameters $\bm{\mu}$ and $\bm{x_0}$.
\noindent

\bigskip

\noindent\textbf{Input: (\textit{Toy example})}
\[%
  \begin{array}
    [c]{lll}%
    \Sigma & : & \left\{
    \begin{array}
      [c]{lll}%
      x^{\prime}        & = & -\mu x \\
      y                 & = & x^2+x  \\
      x\left(  0\right) & = & x_{0}  %
    \end{array}
    \right.                     \\
    D      & : & 
    \left(0.000, 2.000\right),
    \left(0.333,1.563\right), 
    \left(0.666,1.229\right),
    \left(1.000, 0.974\right) 
  \end{array}
\]
\noindent\textbf{Output:}  $(\mu,x_0) \;\approx\; (0.499,1.000)$
\medskip

\noindent The result above is  good because we have ``simulated'' the measured output values~$D$ by numerically
solving the model~$\Sigma$ for $\mu=0.500$ and $x_0=1.000$.

\bigskip

\noindent To solve the toy example problem using our software, enter the following lines into Julia console:

\medskip

\label{app:toy_example}
{\footnotesize
\begin{lstlisting}
using ParameterEstimation
using ModelingToolkit

# --- Model
@parameters mu
@variables  t x(t) y(t)
D = Differential(t)
@named Sigma = ODESystem([D(x) ~ -mu * x],t,[x],[mu])
outs = [y ~ x^2 + x]

# --- Data
data = Dict(
  "t"     => [0.000, 0.333, 0.666, 1.000],
  x^2 + x => [2.000, 1.563, 1.229, 0.974])

# --- Run
res = estimate(Sigma, outs, data);
\end{lstlisting}
}
\smallskip

\noindent You should see the following output:

\smallskip

{\footnotesize
\begin{lstlisting}
Parameter(s)        : mu   = 0.499
Initial Condition(s): x(t) = 1.000
\end{lstlisting}
}

\section{The proposed Approach}
\label{sec:features}
We will briefly describe the proposed approach by first  giving a flow chart of the steps.

\bigskip

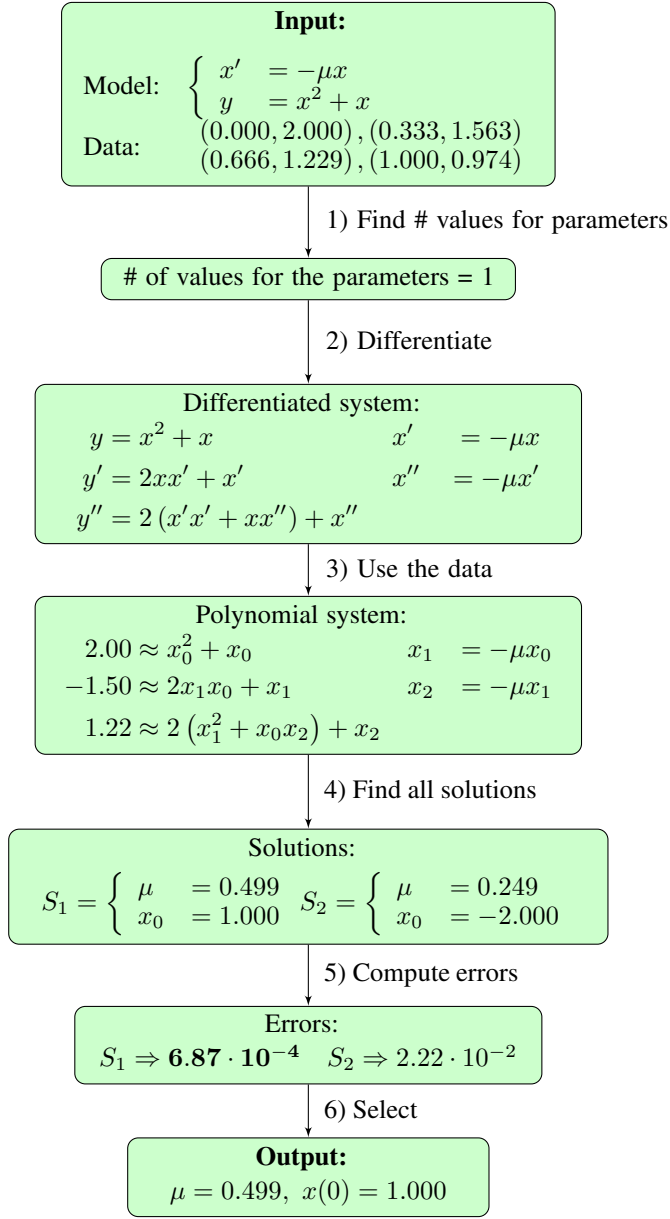
\tikzstyle{block} =
[rectangle,
draw,
fill=green!20,
text width=10em,
text centered,
rounded corners,
minimum height=0em]
\tikzstyle{line} = [draw, -latex']
\tikzstyle{sline} = [draw]
\tikzstyle{dummy} = [circle, minimum size=0pt, inner sep = 0pt]
\begin{tikzpicture}[auto]
    \node [block,  text width=18em] (diffeq) {
      {\bf Input:}
      \[\begin{array}{ll}
          \text{Model:} &
          \left\{\begin{array}{ll}
                   x' &=  -\mu x \\
                   y  &=  x^2+x
                 \end{array} \right. \\

          \text{Data:}  &
          \begin{array}{l}
            \left(0.000, 2.000\right), \left(0.333,1.563\right) \\
            \left(0.666, 1.229\right), \left(1.000, 0.974\right)
          \end{array}
        \end{array}\]
    };

    \node [block, below of = diffeq, text width=15em, node distance = 7em] (id){
      \# of values for the parameters = 1
     };

    \node [block, below of = id, text width=20em, node distance = 7em] (algeq){
      Differentiated system:
      \vspace{-0.5em}
      \begin{align*}
        y                & = x^2+x      &&       x^{\prime}  & =  -\mu x    \\
        y^{\prime}       & = 2xx^{\prime}+x^{\prime} &&  x^{\prime\prime} & =  -\mu x^{\prime} \\
        y^{\prime\prime} & =  2\left(x' x' + xx^{\prime\prime}\right) + x^{\prime\prime} 
      \end{align*}
    };

    \node [block, below  of = algeq, text width=20em,node distance =8em] (qalgebra){
      Polynomial system:
      \vspace{-0.5em}
      \begin{align*}
        2.00  & \approx  x_0^2+x_0      && x_1   & =  -\mu x_0  \\
        -1.50 & \approx  2x_1x_0 + x_1  && x_2   & =  -\mu x_1  \\
        1.22  & \approx  2\left(x_1^2 + x_0x_2\right)+x_2 
      \end{align*}
    };

    \node [block,  below of = qalgebra, text width=22em, node distance =8em] (qsolving) {
      Solutions:
      \vspace{-0.2em}
        \begin{align*}
          S_1 =\left\{
          \begin{array}{ll}
            \mu & = 0.499 \\
            x_0 & = 1.000
          \end{array}
          \right. 
          S_2 =\left\{
          \begin{array}{ll}
            \mu & = 0.249  \\
            x_0 & = -2.000
          \end{array}
          \right.
        \end{align*}
    };

    \node [block,  below of = qsolving, text width=17em, node distance = 6em] (qsolving2){
      Errors:
      \vspace{-0.5em}
      \[
        \begin{array}{ll}
          S_1 \Rightarrow {\bf 6.87\cdot 10^{-4}} &
          S_2 \Rightarrow  2.22\cdot 10^{-2}
        \end{array}
      \]
    };

    \node [block, below of = qsolving2, text width=13em, node distance = 5em] (qanalysis){
      {\bf Output:}
      \vspace{-0.5em}
      \begin{align*}\mu=0.499,~x(0)=1.000 \end{align*}
    };

    (diffeq)--
    node [midway] {}
    (qanalysis);
    \path [line]
    (diffeq)--
    node [text width=14em] {~1) Find \#  values for parameters }
    (id);
    \path [line]
    (id)--
    node [text width=10em] {~2) Differentiate}
    (algeq);
    \path [line]
    (algeq)--
    node [text width=10em] {~3) Use the data }
    (qalgebra);
    \path [line]
    (qalgebra)--
    node [text width==10em] {~4) Find all solutions}
    (qsolving);
    \path [line]
    (qsolving)--
    node [text width==10em] {~5) Compute errors}
    (qsolving2);
    \path [line]
    (qsolving2)--
    node [text width==10em] {~6) Select}
    (qanalysis);

  \end{tikzpicture}

Now we elaborate on each step in the above flow chart.
\begin{enumerate}
\item \label{iden} Use the model~$\Sigma$ to find the number of values for parameters (see \cite{hong_sian_2019,hong_global_2020} for details, cf. \cite{bates2019identifiability}).

        \textit{Toy example:}
        
        There is only one value for the parameters.
  \item 
        \label{step:diff} Differentiate the model~$\Sigma$ sufficiently many times (see \cite{hong_sian_2019,hong_global_2020} for details).
        
       \smallskip
        \textit{Toy example:}%
        \[
          \left\{
          \begin{array}
            [c]{lll}%
            y                & = & x^2+x                                                                     \\
            y^{\prime}       & = & 2xx^{\prime}+x^{\prime}                                                   \\
            y^{\prime\prime} & = & 2\left(x^{\prime}x^{\prime} + xx^{\prime\prime}\right) + x^{\prime\prime} \\
            x^{\prime}       & = & -\mu x                                                                    \\
            x^{\prime\prime} & = & -\mu x^{\prime}                                                           %
          \end{array}
          \right.
        \]
        \item\label{step:3} Use the data~$D$ to derive a numerical overdetermined polynomial system from the above equations.

        \begin{enumerate}
          \item Approximate the output functions $y$ by interpolating the data $D$ into   rational functions~$g$.

                \smallskip
                \textit{Toy example:}%
                \[
                  \begin{array}{ll}
                    y & \approx\; g\;= \;\frac{0.58t^2 - 3.11t + 6.82}{t + 3.41} \\
                  \end{array}
                \]

          \item Approximate the system, obtained from Step \ref{step:diff}), by replacing~$y$ with $g$.
                
                \smallskip
                \textit{Toy example:}%
                \[
                  \left\{
                  \begin{array}
                    [c]{rll}%
                    \frac{0.58t^2 - 3.11t + 6.82}{t + 3.41}                   & \approx & x^2+x                                                                    \\
                    \frac{0.58 t^2 + 3.96 t - 17.41}{\left(t + 3.41\right)^2} & \approx & 2xx^{\prime} + x^{\prime}                                                \\
                    \frac{48.32}{\left(t + 3.41\right)^3}                     & \approx & 2\left(x^{\prime}x^{\prime} + xx^{\prime\prime}\right) +x^{\prime\prime} \\
                    x^{\prime}                                                & =       & -\mu x                                                                   \\
                    x^{\prime\prime}                                          & =       & -\mu x^{\prime}                                                          %
                  \end{array}
                  \right.
                \]

          \item Evaluate the above system at $t=0$.

               \smallskip
                \textit{Toy example:}%
                \[
                  \left\{
                  \begin{array}
                    [c]{rll}%
                    2.00  & \approx & x_0^2+x_0,                        \\
                    -1.50 & \approx & 2x_1x_0 + x_1,                    \\
                    1.22  & \approx & 2\left(x_1x_1 + x_0x_2\right)+x_2 \\
                    x_1   & =       & -\mu x_0                          \\
                    x_2   & =       & -\mu x_1
                  \end{array}
                  \right.
                \]
                where a shorthand notation $x_{i}=x^{\left(i\right)  }\left(0\right)  $ is used. Note
                that the system consists of~$5$~equations in~$4~$unknowns: $\mu, x_{0}, x_{1}, x_{2}$.

        \end{enumerate}

        \item\label{step:4} Find all the solutions of the over-determined system of polynomial equations.
        For this, first we make the system square.
        Then we find all the real solutions of the square system (for instance, using homotopy continuation method, see \cite{HomotopyContinuation.jl}).

       \smallskip
        \textit{Toy example:}
        \vskip -2em
        \[\begin{cases}
            (\mu ,x_0) \;\approx\; (0.499,\;\;\;1.000) \\
            (\mu ,x_0) \;\approx\; (0.249,-2.000)
          \end{cases}\]

  \item \label{step:5} Compute errors of the solutions.

        For each solution $\left(\widehat{\mu},\widehat{x}_{0}\right)$
        \begin{enumerate}
          \item Compute $\widehat{y}\left(t_{i}\right)$ by solving $\Sigma\left(\widehat{\mu},\widehat{x}_{0}\right)$
                using numerical ODE solver.

          \item Compute error $e$ between the computed values of $\widehat{y}$ and the measured values of $y.$
        \end{enumerate}

        \textit{Toy example:}
        \[\begin{cases}
            (\mu, x_0) \approx (0.499,\;\;\; 1.0000)\Rightarrow e=6.87\cdot 10^{-4} \\
            (\mu, x_0) \approx (0.249,-2.0000)\Rightarrow e=2.22\cdot 10^{-2}
          \end{cases}\]

  \item \label{step:6} Select $k$ solutions with smallest error, where~$k$ is the number of parameter values from Step~\ref{iden}).  
 If the user has a priori information about parameter ranges, we can first filter the solutions before choosing the $k$ best solutions.

\smallskip
 \textit{Toy example:}  $(\mu,x_0) \;\approx\; (0.499,1.0000)$
\end{enumerate}

\section{Implementation details}

\subsection*{Steps~\ref{iden}) and~\ref{step:diff})}
The two steps are are done by calling
SIAN \cite{hong_sian_2019,hong_global_2020}.

\subsection*{Step~\ref{step:3})}
We perform interpolation of the dataset~\(D\).
In this interpolation, we have considerable freedom in choosing the interpolation method, with the following desired properties:
\begin{enumerate}
\item easily calculated analytic derivatives, which are as well interpolated as the function itself
\item the ability to fit a wide variety of shapes, including asymptotic behavior, decay to 0, and periodicity, 
\item a preference for parsimonious models, and
\item numerical stability.
\end{enumerate} Our software supports a variety of modern interpolation methods, via \pn{BaryRational.jl} and other interpolation packages.  Of note, we support Fourier extension, Floater-Hormann rational interpolation \cite{FloaterHormann}, as well as the  ``AAA" algorithm introduced in \cite{AAAAlgorithm}, which uses a barycentric rational representation adapted to input data.  Experimental data show that AAA rational interpolation interpolates smooth functions in such a way that,  for a wide variety of data outputs from differential equations, even higher order derivatives are well interpolated.  In our software, we use AAA as a default interpolation scheme, but allow iterating over multiple schemes in a parallel fashion to select the best one.

Once interpolation is done, we apply high-order automatic
differentiation provided by the \pn{TaylorSeries.jl}
\cite{benet2019taylorseries}  and \pn {ForwardDiff.jl} packages to the interpolated function to estimate derivative values  of higher order.

We substitute the estimated values of the derivatives of the output $y$  into the system obtained in Step~\ref{step:diff}). This typically results in  an over-determined system of polynomial equations.

\subsection*{Step~\ref{step:4})}
Since the obtained polynomial system is typically overdetermined, the next step is to square the system carefully.
This is achieved by collecting equations one by one, given that
a newly added equation increases the rank of the Jacobian of the system.
If, for a given equation, the rank is unchanged, the equation is not
added. Then we find all solutions of the squared system. By default, we are using homotopy-based solving from  \pn{HomotopyContinuation.jl}. We also support \pn{MSolve} \cite{msolve} via Oscar algebra system in Julia \cite{OSCAR,OSCAR-book}. This method finds all solutions with theoretical guarantees. However, it has been less stable for some of the systems we encountered. All results presented here use homotopy-based solving, the default option.

\subsection*{Steps~\ref{step:5}) and ~\ref{step:6})}
For each obtained solution (parameter estimates), we compute the data by numerically solving the model. Then we determine errors between the computed data and the given (measured) data.
We keep the estimates with minimal errors. Such estimates
may exist for different interpolation schemes; we keep the ones with minimal error across all schemes.

\subsection*{Extra features}
\noindent We would like to emphasize the following extra features:

\begin{itemize}\itemsep=0.5em
  \item The iteration over interpolation schemes and underlying
        polynomial solving via \pn{HomotopyContinuation.jl} \cite{HomotopyContinuation.jl}
        are parallelizable and can take advantage of a
        multicore (or multithreaded) computing environment.
  \item \pn{ParameterEstimation.jl} provides identifiability interface to
        SIAN \cite{hong_sian_2019,hong_global_2020}. One can assess local and global identifiability of the
        parameters and initial conditions before estimation.
\end{itemize}

\section*{Current Limitations}
\noindent 
We list a few limitations of the current  preliminary implementation:

\begin{itemize}\itemsep=0.5em
  \item {As a two-stage approach, the issues common to this class of approaches apply. The data must be time series data which is sufficiently dense for the rational polynomial fitting to give accurate derivative approximations.}
  \item Our software supports rational ODEs models (the right-hand sides are rational functions), while software such as AMIGO2 and IQM also support ODE models that are not necessarily rational.
  \item {We use SIAN that is based on a Monte-Carlo algorithm that returns the bound on number of solutions with high probability, though detailed analysis is still ongoing.}
  \item {The high order derivative approximations via rational polynomials may suffer from errors due to noise or rounding. We leave the extension of the technique to usage with robust derivative approximations \cite{li2002estimation,chartrand2011numerical} as a topic of further study.}
\end{itemize}


\section{Performance}
\medskip

\begin{table*}[!h]
\centering

{\large \bf Median of Relative Errors in \% }

\medskip

\resizebox{0.99\textwidth}{!}{
\begin{tabular}{||c|l||r|r|r|r|r|r|r|r|r|c||}
\hline
\multicolumn{2}{||l||}{\multirow{2}{*}{Software}}&
\multicolumn{3}{c|}{\multirow{2}{*}{IQM}} &
\multicolumn{3}{c|}{\multirow{2}{*}{SciML}} &
\multicolumn{3}{c|}{\multirow{2}{*}{AMIGO2}} &
\multicolumn{1}{c||}{Parameter}\\

\multicolumn{2}{||l||}{} & 
\multicolumn{1}{c}{}& \multicolumn{1}{c}{} &\multicolumn{1}{c|}{}& 
\multicolumn{1}{c}{}& \multicolumn{1}{c}{} &\multicolumn{1}{c|}{}& 
\multicolumn{1}{c}{}& \multicolumn{1}{c}{} &\multicolumn{1}{c|}{}&
\multicolumn{1}{c||}{Estimation.jl}\\
\hline

\multicolumn{2}{||l||}{Search Range} 
 &
\multicolumn{1}{c|}{[0,1]} & \multicolumn{1}{c|}{[0,2]} & \multicolumn{1}{c|}{[0,3]} & 
\multicolumn{1}{c|}{[0,1]} & \multicolumn{1}{c|}{[0,2]} & \multicolumn{1}{c|}{[0,3]} & 
\multicolumn{1}{c|}{[0,1]} & \multicolumn{1}{c|}{[0,2]} & \multicolumn{1}{c|}{[0,3]} & 
\multicolumn{1}{c||}{Any} \\
\hline\hline

& {Harmonic, \cref{simple}}
& 64.2	& 63.0	& 65.1	
& 0.0	& 0.0	& 0.0	
& 0.0	& 0.0	& 0.0	
& 0.0 \\
\cline{2-12}

& {Van der Pol, \cref{vdp}}
& 0.0	& 0.0	& 0.0	
& 0.0	& 0.0	& 0.0	
& 0.0	& 0.0	& 0.0	
& 0.0 \\
\cline{2-12}

& {FitzHugh-Nagumo, \cref{fhn}}
& 68.8	& 139.2	& 159.7	
& 0.6	& 0.2	& 0.3	
& 0.0	& 0.0	& 0.0	
& 0.0 \\
\cline{2-12}

& {HIV, \cref{hiv}}
& 75.5	& 89.7	& 133.8	
& 2.0	& 7.9	& 57.8	
& 0.0	& 0.0	& 0.0	
& 0.0 \\
\cline{2-12}

& {Mammillary 3, \cref{daisy_mamil3}}
& 74.3	& 88.1	& 85.6	
& 5.6	& 12.8	& 11.8	
& 0.0	& 0.0	& 0.0	
& 0.0 \\
\cline{2-12}

& {Lotka-Volterra, \cref{lv}}
& 71.7	& 76.5	& 76.5	
& 22.1	& 55.0	& 75.6	
& 0.0	& 0.0	& 1.0	
& 0.0 \\
\cline{2-12}

& {Crauste, \cref{crauste}}
& 85.3	& 109.3	& 139.8	
& 26.5	& 67.2	& 144.8	
& 0.0	& 77.6	& 24.5	
& 0.0 \\
\Xcline{2-12}{2\arrayrulewidth}

& {Biohydrogenation, \cref{bioh}}
& 81.2	& 113.4	& 78.9	
& 53.9	& 110.6	& 284.6	
& 0.0	& 0.0	& 22.2	
& 0.0 \\
\cline{2-12}

& {Mammillary 4, \cref{daisy_mamil4}}
& 76.0	& 88.4	& 105.1	
& 46.9	& 61.1	& 97.1	
& 0.0	& 53.8	& 51.9	
& 0.0 \\
\cline{2-12}

\parbox[t]{2mm}{\multirow{-10}{*}{\rotatebox[origin=c]{90}{Models}}}
& {SEIR, \cref{seir}}
& 101.7	& 173.0	& 258.1	
& 17.1	& 49.6	& 105.4	
& 17.9	& 19.8	& 30.8	
& 0.0 \\
\hline
\end{tabular}
}

\bigskip

{\large \bf Mean of Relative Errors in \% }

\medskip

\resizebox{0.99\textwidth}{!}{
\begin{tabular}{||c|l||r|r|r|r|r|r|r|r|r|c||}
\hline
\multicolumn{2}{||l||}{\multirow{2}{*}{Software}}&
\multicolumn{3}{c|}{\multirow{2}{*}{IQM}} &
\multicolumn{3}{c|}{\multirow{2}{*}{SciML}} &
\multicolumn{3}{c|}{\multirow{2}{*}{AMIGO2}} &
\multicolumn{1}{c||}{Parameter}\\

\multicolumn{2}{||l||}{} & 
\multicolumn{1}{c}{}& \multicolumn{1}{c}{} &\multicolumn{1}{c|}{}& 
\multicolumn{1}{c}{}& \multicolumn{1}{c}{} &\multicolumn{1}{c|}{}& 
\multicolumn{1}{c}{}& \multicolumn{1}{c}{} &\multicolumn{1}{c|}{}&
\multicolumn{1}{c||}{Estimation.jl}\\
\hline

\multicolumn{2}{||l||}{Search Range} 
 &
\multicolumn{1}{c|}{[0,1]} & \multicolumn{1}{c|}{[0,2]} & \multicolumn{1}{c|}{[0,3]} & 
\multicolumn{1}{c|}{[0,1]} & \multicolumn{1}{c|}{[0,2]} & \multicolumn{1}{c|}{[0,3]} & 
\multicolumn{1}{c|}{[0,1]} & \multicolumn{1}{c|}{[0,2]} & \multicolumn{1}{c|}{[0,3]} & 
\multicolumn{1}{c||}{Any} \\
\hline\hline

& {Harmonic, \cref{simple}}
& 66.3	& 68.7	& 101.9	
& 0.0	& 0.0	& 0.0	
& 0.0	& 0.0	& 0.0	
& 0.0 \\
\cline{2-12}

& {Van der Pol, \cref{vdp}}
& 7.5	& 0.0	& 9.6	
& 0.0	& 0.0	& 0.0	
& 0.0	& 0.0	& 0.0	
& 0.0 \\
\cline{2-12}

& {FitzHugh-Nagumo, \cref{fhn}}
& 82.4	& 131.1	& 210.3	
& 8.3	& 15.9	& 20.4	
& 0.0	& 0.0	& 0.2	
& 0.0 \\
\cline{2-12}

& {HIV, \cref{hiv}}
& 85.6	& 104.2	& 144.7	
& 14.8	& 36.9	& 102.9	
& 18.5	& 43.2	& 28.4	
& 0.0 \\
\cline{2-12}

& {Mammillary 3, \cref{daisy_mamil3}}
& 76.1	& 104.3	& 128.2	
& 11.9	& 21.5	& 24.7	
& 0.0	& 0.0	& 0.0	
& 0.0 \\
\cline{2-12}

& {Lotka-Volterra, \cref{lv}}
& 72.2	& 75.4	& 75.4	
& 54.7	& 76.9	& 111.7	
& 13.6	& 112.6	& 29.7	
& 0.0 \\
\cline{2-12}

& {Crauste, \cref{crauste}}
& 99.6	& 125.2	& 179.7	
& 47.3	& 72.9	& 226.4	
& 2.6	& 105.7	& 51.7	
& 0.0 \\

\Xcline{2-12}{2\arrayrulewidth}

& {Biohydrogenation, \cref{bioh}}
& 93.7	& 184.0	& 130.3	
& 77.7	& 151.1	& 306.6	
& 17.8	& 13.5	& 28.3	
& 0.0 \\
\cline{2-12}

& {Mammillary 4, \cref{daisy_mamil4}}
& 94.1	& 109.6	& 105.3	
& 66.2	& 67.7	& 118.8	
& 29.8	& 56.7	& 59.0	
& 0.1 \\
\cline{2-12}

\parbox[t]{2mm}{\multirow{-10}{*}{\rotatebox[origin=c]{90}{Models}}}
& {SEIR, \cref{seir}}
& 132.3	& 230.8	& 338.9	
& 18.4	& 60.6	& 118.1	
& 25.9	& 40.7	& 50.8	
& 0.0 \\
\hline
\end{tabular}
}

\bigskip

\caption{ The above two tables show the median and the mean of the Root Mean Square Relative Errors (RMSRE) over~10 randomly generated data sets.
For the details on how the data sets are generated and how the errors are computed, see Subsections~\ref{sec:settings},~\ref{sec:data} and~\ref{sec:err}. The last three models are grouped separately from the rest to indicate that the model has parameters that are only locally identifiable.} 
\label{tab:benchmarking_err}
\end{table*}

In this section, we study the performance (in particular robustness) of several software packages on a collection of benchmarks models listed in \cref{sec:illustrate}. 

\subsection{Software Packages}

\noindent We ran the following software packages:

\begin{itemize} \itemsep=0.5em
  \item IQM \cite{SBtoolbox}

        It uses a single shooting approach with a local derivative-free optimizer (Nelder-Mead, as specified by \pn{symplexIQM}).
        
  \item SciML \cite{DifferentialEquations.jl-2017,rackauckas2020universal}
  
        It is intentionally a naive single shooting implementation which serves as a baseline for one of the most common parameter estimation approaches. While such a package can easily be improved using methods like multiple shooting and global optimizers, this setup demonstrates the expected results from the standard simple approach. A polyalgorithm mixing ADAM with BFGS is used to demonstrate the results of a common gradient-based optimization scheme. 
        
  \item AMIGO2 \cite{AMIGO2}
  
        It uses a multi-start global optimization approach.  This approach can always be tuned to achieve correct results by choosing sufficient bounds/tolerances and population points, though not all of these values are typically known in advance. 
        
  \item \pn{ParameterEstimation.jl} (this work)
\end{itemize}

\noindent We could not run the following  software packages:
Data2Dynamics \cite{Data2Dynamics}, COPASI \cite{COPASI}, and IQR\footnote{https://iqrtools.intiquan.com/}/IQDesktop\footnote{https://iqdesktop.intiquan.com/book/}
because of technical difficulties in installation.

\subsection{Settings used for Software Packages}
\label{sec:settings} 
IQM, SciML and AMIGO2 require search ranges or initial guesses for the parameters and initial conditions or certain additional settings to be specified by the user that pertain specifically to the optimization task performed by each software package, for instance,
maximal \# of iterations for gradient-based optimization,
specific loss/error functions to be optimized.

ParameterEstimation.jl is built on top of existing algorithms for solving polynomial systems  and interpolating data points. Their implementations involve some settings.

For studying performance of the software systems, we used the following settings:
\begin{itemize}\itemsep=0.3em
\item 
For IQM, we set the maximal number of function
evaluations at 200,000 and  used a \pn{simplexIQM} solver. 
\item 
For SciML, we set the number of function evaluation at 200,000 and set the learning rate at 0.01.
\item
For AMIGO2, we  set the maximal number of function
evaluations at 200,000, set the IVP solver integration tolerances at $10^{-13}$ and set the solver tolerances to $10^{-13}$
and used a nonlinear least squares solver, which meets or exceeds the settings suggested by Julio Banga\footnote{Private communication, May-June 2023}.
Due to the non-deterministic nature of choosing starting points in the global optimization used by AMIGO2, multiple runs with the same example and settings can produce different estimates (possibly with larger or smaller error). We only ran it once and reported the results. 

\item  For ParameterEstimation.jl, default options were used for most settings.  The default polynomial solver is HomotopyContinuation.jl, with a tolerance of $10^{-12}$.   
The default setting is to use a fixed set of interpolation schemes including AAA, Floater-Hormann of orders (3,6,8), and Fourier interpolation (though, for all of the models in this paper, AAA interpolation alone is sufficient.)
\end{itemize}

\subsection{Test Data Generation}
\label{sec:data} 

For all models and all software packages, the test data were generated uniformly as follows.
We chose 21 time points equally spaced over $[-0.5,0.5]$ and also chose the reference values of the parameters and initial conditions randomly from the interval $[0.1,0.9]$  using the {random.uniform} function in Python's {numpy} package. 
For reproducibility, we used {random.seed(0)} at the start of choosing the 10 data sets for each system. In case of errors (e.g., due to singularity), the set was discarded and the next set of random values were generated instead.
These uniform random choices were made for the sake of studying/comparing the robustness of the approaches in isolation, independent of the physical systems underlying the models. The random values were at least 0.1 chosen away from 0.0 and 1.0 to avoid potential degeneracy. 

For each model and for each software package, the input script (containing the model description, the generated data and settings) was  generated automatically.

\subsection{Estimation Error}
\label{sec:err}
To evaluate the quality of the estimation, for each data set, we computed the \emph{Root Mean Square Relative Error} defined as
\[
  RMSRE = \sqrt{\frac{1}{\#P}\sum_{\substack{p} \in P} \left( \frac{p_{true}-p_{estimated}}{p_{true}}\right)^2}  \times 100,
\]
where $P$ denotes the  identifiable\footnote{ For the Biohydrogenation model, we excluded the errors from the estimates for the initial value of $x_7$, which is not identifiable.} set of the system parameters and the initial values of the state variables in the input system.
The appendix provides RMSRE for each model, search interval, and random data set.
Table~\ref{tab:benchmarking_err} summarizes them using median and mean.

\subsection{Discussion}
The inspection of the test results in Table~\ref{tab:benchmarking_err}  and Appendix shows that the proposed approach outperforms the others with regard to estimation error in all but one instance.
Furthermore, the proposed approach is robust in the following aspects:

\smallskip

\noindent {\bf Search Interval:}
See Table~\ref{tab:benchmarking_err}. Note that IQM, SciML and AMIGO2 (based on shooting approaches) require the user to specify a search interval. Note also that their relative errors are \emph{dependent} on search interval, sometimes even \emph{non-monotonically}. 
For an example, see the relative error of AIMOG2 on the Crauste Model. If the upper bound of the search interval is increased from 1.0 to 2.0, the median (mean) of the relative error is increased from 0.0\% (2.6\%) to 77.6\% (105.7\%). However, if the upper bound is increased to 3.0, the median (mean) of the relative error is decreased to 24.5\% (51.7\%). See Random Data 3, 5, 9 and 10 in the Crauste table of Appendix for specific instances of this behavior.  
This non-monotonic behavior could add difficulty to the users in specifying a suitable search interval.

The proposed approach's relative error is, by design, \mbox{\emph{independent}} of search interval, because it does not require/use search interval.

\smallskip

\noindent {\bf Parameter values:} 
See the tables in Appendix.  Note that the relative errors of IQM, SciML and AMIGO2 are \emph{dependent} on the values of the parameters (randomly sampled from [0.1, 0.9]), sometime drastically. 
For an example, see the Crauste table. AMIGO2 has essentially no error for some random values of parameters but huge error for the others.

The proposed approach's relative error is practically \mbox{\emph{independent}}  of the values of the parameters.
\smallskip
 
\noindent {\bf Multiple solutions:}
See Table~\ref{tab:benchmarking_err}. The first seven models are globally identifiable (hence with unique solution for the parameters). The last three models contain parameters that are only locally identifiable (hence with multiple solutions for the parameter values).  Note that the relative errors of IQM, SciML and AMIGO2 are significantly larger for locally identifiable models.   
It is because the optimization-based approaches may miss an intended value of a parameter.

The proposed approach's relative error is practically \mbox{\emph{independent}}  of identifiability, as long as it is at least locally identifiable.
In fact, it finds \emph{all} the solutions, making it easier for the users to quickly understand the full landscape of solutions.

\section{List of Differential Models}
\label{sec:illustrate}
In this section, we present the ODE systems used in the benchmarks.

\subsection*{Harmonic Oscillator Model}
It is a toy model for harmonic oscillators without damping.
\begin{align}
   & \begin{cases}
       \dot{x_1} = -a\,x_2 \\
       \dot{x_2} = \frac{1}{b}\,x_1
     \end{cases}
  \label{simple}
  \\
   & \begin{cases}
       y_1 = x_1,\\
       y_2 = x_2
     \end{cases}
  \label{simple:outs}
\end{align}

\subsection*{Van der Pol Oscillator Model}
It models harmonic oscillators with nonlinear damping
\cite{vanderPol1920}, 
which was introduced to study oscillations in electronic circuits, 
but was found to be useful for various other phenomena such as
vocal fold~\cite{Lucero2013}.

\begin{align}
   & \begin{cases}
       \dot{x_1} = a x_2 \\
       \dot{x_2} = -x_1 - b(x_1^2 -1)x_2
     \end{cases}
  \label{vdp}
  \\
   & \begin{cases}
       y_1 = x_1,  \\
       y_2 = x_2
     \end{cases}
  \label{vdp:outs}
\end{align}

\subsection*{FitzHugh-Nagumo Model}
It models (two-dimensional simplification of) spike generation in squid giant axons \cite{fitzhugh1961impulses,nagumo1962active}.
\begin{align}
   & \begin{cases}
       \dot{V} = g\,\left(V - \frac{V^3}{3} + R\right) \\
       \dot{R} = \frac{1}{g}\,\left(V - a + b\,R\right)
     \end{cases}
  \label{fhn}
  \\
   & \begin{cases}
       y_1 = V
     \end{cases}
  \label{fhn:outputs}
\end{align}

\subsection*{HIV Dynamics Model}
It models HIV
infection dynamics during
interaction with the immune system
over the course of various treatments. It is based on 
~\cite[equations~(6)]{WN2002} and the paragraph that follows.
\begin{align}
  \label{hiv}
   & \begin{cases}
       \dot{x} = \lambda - d\,x - \beta\,x\,v   \\
       \dot{y} = \beta\,x\,v - a\,y             \\
       \dot{v} = k\,y - u\,v                    \\
       \dot{w} = c\,x\,y\,w - c\,q\,y\,w - b\,w \\
       \dot{z} = c\,q\,y\,w - h\,z
     \end{cases}
  \\
   & \begin{cases}
       y_1 = w,\\ y_2 = z, \\
       y_3 = x, \\ y_4 = y + v
     \end{cases}
  \label{hiv:outputs}
\end{align}

\subsection*{Mammillary 3-compartment  Model}
It models pharmacokinetic processes, such as protein metabolism in organism
~\cite{benet1972general,hui2005adaptive}.
\begin{align}{}
   & \begin{cases}
       \dot{x_1} = -(a_{21} + a_{31} + a_{01})\,x_1 + a_{12}\,x_2 + a_{13}\,x_3 \\
       \dot{x_2} = a_{21}\,x_1 - a_{12}\,x_2                                    \\
       \dot{x_3} = a_{31}\,x_1 - a_{13}\,x_3
     \end{cases}
  \label{daisy_mamil3}
  \\
   & \begin{cases}
       y_1 = x_1, \\
       y_2 = x_2
     \end{cases}
  \label{daisy_mamil3:outputs}
\end{align}

\subsection*{Lotka-Volterra Model}
It models interactions between two species (a predator and a prey) in an ecosystem ~\cite{lotka1920analytical,volterra1928variations}.
\begin{align}
   & \begin{cases}
       \dot{r} = k_1\,r - k_2\,r\,w, \\
       \dot{w} = k_2\,r\,w - k_3\,w
     \end{cases}
  \label{lv}
  \\
   & \begin{cases}
       y_1=r
     \end{cases}
  \label{lv:outputs}
\end{align}

\subsection*{Crauste Model}
It models the behavior of CD8 T-cells. It is 
based on the Crauste system \cite[Section 4]{CrausteSource2} originally described in \cite{CrausteSource1}.
\begin{align}
   & \begin{cases}
       \dot{N} = -\mu_N\,N - \delta_{NE}\,N\,P                                     \\
       \dot{E} = \delta_{NE}\,N\,P - \mu_{EE}\,E^2 - \delta_{EL}\,E + \rho_E\,E\,P \\
       \dot{S} = \delta_{EL}\,S - S\,\delta_{LM} - \mu_{LL}\,S^2 - \mu_{LE}\,E\,S  \\
       \dot{M} = \delta_{LM}\,S - \mu_M\,M                                         \\
       \dot{P} = \rho_P\,P^2 - \mu_P\,P - \mu_{PE}\,E\,P -\mu_{PL}\,S\,P
     \end{cases}
  \label{crauste}
  \\
   & \begin{cases}
       y_1 = N,\\ y_2 = E,\\
       y_3 = S + M,\\ y_4 = P
     \end{cases}
  \label{crauste:outputs}
\end{align}

\subsection*{Biohydrogenation Model}
 It models the kinetic processes involved in biohydrogenation of non-esterified fatty acids ~\cite{BiohydrogenationSource1, BiohydrogenationSource2}.

\begin{align}
   & \begin{cases}
       \dot{x_4} = - \frac{k5 \, x_4 }{k_6 + x_4} \\
       \dot{x_5} = \frac{k_5 \, x_4 }{k_6 + x_4} - \frac{k_7 \, x_5}{k_8 + x_5 + x_6} \\
       \dot{x_6} = \frac{k_7 \, x_5 }{k_8 + x_5 + x_6} - \frac{k_9 \, x_6 \, (k_{10} - x_6) }{k_{10}} \\
       \dot{x_7} = \frac{k_9 \, x_6 \, (k_{10} - x_6) }{k_{10}}
     \end{cases}
  \label{bioh}
  \\
   & \begin{cases}
       y_1 =  x_4, \\
       y_2 =  x_5 
     \end{cases}
  \label{bioh:outputs}
\end{align}

\subsection*{Mammillary 4-Compartment Model}

It comes from examples used in the analysis of DAISY
\cite{DAISY} identifiability software.
It represents a
4-compartment system.
\begin{align}
   & \begin{cases}
       \dot{x_1}  =  -k_{01} \, x_1 + k_{12} \, x_2 + k_{13}\, x_3 +        \\
       \hspace{3.5em}k_{14}x_4 - k_{21}\, x_1 - k_{31}\, x_1 - k_{41}\, x_1 \\
       \dot{x_2}  =  -k_{12} \, x_2 + k_{21} \, x_1                         \\
       \dot{x_3}  =  -k_{13} \, x_3 + k_{31} \, x_1                         \\
       \dot{x_4}  =  -k_{14} \, x_4 + k_{41} \, x_1                         \\
     \end{cases}
  \label{daisy_mamil4}
  \\
   & \begin{cases}
       y_1 =  x_1,\\
       y_2 =  x_2,\\
       y_3 =  x_3+x_4
     \end{cases}
  \label{daisy_mamil4:outputs}
\end{align}

\subsection*{SEIR Model}
 It models an epidemic that involves
compartments related to disease progression stages with prevalence observations~\cite{SEIRSource}.
\vskip-1em
\begin{align}
   & \begin{cases}
    \dot{S} = -b \, S \, I / N\\
    \dot{E} = b \, S \, I / N - \nu \, E\\
    \dot{I} = \nu \, E - a \, I\\
    \dot{N} = 0
     \end{cases}
  \label{seir}
  \\
   & \begin{cases}
       y_1 = I,\\
       y_2 = N
     \end{cases}
  \label{seir:outputs}
\end{align}

\section*{Acknowledgements}
We, the authors, are grateful 
to Julio Banga and Eva Balsa-Canto for extensive feedback on the use of AMIGO2,
to Henning Schmidt, developer of IQRTools/IQDesktop for numerous discussions on installation of the software,
to Gleb Pogudin for helpful feedback on the proposed algorithm and information on other parameter estimation approaches,
to Xinglong Zhou and Jianing Qi for initial explorations of the role of polynomial interpolation,
and finally to the CCiS at Queens College and CIMS NYU for the computational resources.

This work was supported by the National Science Foundation
  [DMS-1760448, CCF-1708884, CCF-2212460, CCF-2212461, CCF-2212462, CCF-1813340], City University of New York [PSC-CUNY \#65605-00 53, This work was supported by the National Science Foundation
    [DMS-1760448, CCF-1708884, CCF-2212460, CCF-2212461, CCF-2212462]
and City University of New York [PSC-CUNY \#65605-00 53, \#66551-00 54], and DARPA [HR00112290091]. This research was, in part, funded by the U.S. Government. The views and
conclusions contained in this document are those of the authors and should
not be interpreted as representing the official policies, either expressed or
implied, of the U.S. Government. Approved for Public Release, Distribution Unlimited.
\bibliography{document.bib,additional_papers.bib}


\bigskip
\bigskip
\centering
\textbf{\large Appendix}
\appendix
{
Harmonic
\vskip 0.5em
 \begin{tabular}{||l||r|r|r|r|r|r|r|r|r|c||}
 \hline
 \multirow{2}{*}{Software}&
 \multicolumn{3}{c|}{\multirow{2}{*}{IQM}} &
 \multicolumn{3}{c|}{\multirow{2}{*}{SciML}} &
 \multicolumn{3}{c|}{\multirow{2}{*}{AMIGO2}} &
 Parameter\\
 & 
 \multicolumn{1}{c}{}& \multicolumn{1}{c}{} &\multicolumn{1}{c|}{}&
 \multicolumn{1}{c}{}& \multicolumn{1}{c}{} &\multicolumn{1}{c|}{}&
 \multicolumn{1}{c}{}& \multicolumn{1}{c}{} &\multicolumn{1}{c|}{}&
 Estimation.jl\\
 \hline
 Search Range  &
 \multicolumn{1}{c|}{[0,1]} & \multicolumn{1}{c|}{[0,2]} & \multicolumn{1}{c|}{[0,3]} &
 \multicolumn{1}{c|}{[0,1]} & \multicolumn{1}{c|}{[0,2]} & \multicolumn{1}{c|}{[0,3]} &
 \multicolumn{1}{c|}{[0,1]} & \multicolumn{1}{c|}{[0,2]} & \multicolumn{1}{c|}{[0,3]} &
 \multicolumn{1}{c||}{Any} \\
 \hline\hline
Random Data 1
& 58.1	& 58.1	& 58.1	
& 0.0	& 0.0	& 0.0	
& 0.0	& 0.0	& 0.0	
& 0.0 \\
\hline
Random Data 2
& 62.5	& 62.5	& 62.5	
& 0.0	& 0.0	& 0.0	
& 0.0	& 0.0	& 0.0	
& 0.0 \\
\hline
Random Data 3
& 54.7	& 54.7	& 327.5	
& 0.0	& 0.0	& 0.0	
& 0.0	& 0.0	& 0.0	
& 0.0 \\
\hline
Random Data 4
& 81.3	& 81.3	& 81.3	
& 0.0	& 0.0	& 0.0	
& 0.0	& 0.0	& 0.0	
& 0.0 \\
\hline
Random Data 5
& 75.8	& 75.8	& 75.8	
& 0.0	& 0.0	& 0.0	
& 0.0	& 0.0	& 0.0	
& 0.0 \\
\hline
Random Data 6
& 64.8	& 0.0	& 0.0	
& 0.0	& 0.0	& 0.0	
& 0.0	& 0.0	& 0.0	
& 0.0 \\
\hline
Random Data 7
& 66.6	& 66.6	& 66.6	
& 0.0	& 0.0	& 0.0	
& 0.0	& 0.0	& 0.0	
& 0.0 \\
\hline
Random Data 8
& 63.5	& 63.5	& 63.5	
& 0.0	& 0.0	& 0.0	
& 0.0	& 0.0	& 0.0	
& 0.0 \\
\hline
Random Data 9
& 79.8	& 168.9	& 227.8	
& 0.0	& 0.0	& 0.0	
& 0.0	& 0.0	& 0.0	
& 0.0 \\
\hline
Random Data 10
& 56.0	& 56.0	& 56.0	
& 0.0	& 0.0	& 0.0	
& 0.0	& 0.0	& 0.0	
& 0.0 \\
\hline
\end{tabular}

\vskip 1em


Van der Pol
\vskip 0.5em
 \begin{tabular}{||l||r|r|r|r|r|r|r|r|r|c||}
 \hline
 \multirow{2}{*}{Software}&
 \multicolumn{3}{c|}{\multirow{2}{*}{IQM}} &
 \multicolumn{3}{c|}{\multirow{2}{*}{SciML}} &
 \multicolumn{3}{c|}{\multirow{2}{*}{AMIGO2}} &
 Parameter\\
 & 
 \multicolumn{1}{c}{}& \multicolumn{1}{c}{} &\multicolumn{1}{c|}{}&
 \multicolumn{1}{c}{}& \multicolumn{1}{c}{} &\multicolumn{1}{c|}{}&
 \multicolumn{1}{c}{}& \multicolumn{1}{c}{} &\multicolumn{1}{c|}{}&
 Estimation.jl\\
 \hline
 Search Range  &
 \multicolumn{1}{c|}{[0,1]} & \multicolumn{1}{c|}{[0,2]} & \multicolumn{1}{c|}{[0,3]} &
 \multicolumn{1}{c|}{[0,1]} & \multicolumn{1}{c|}{[0,2]} & \multicolumn{1}{c|}{[0,3]} &
 \multicolumn{1}{c|}{[0,1]} & \multicolumn{1}{c|}{[0,2]} & \multicolumn{1}{c|}{[0,3]} &
 \multicolumn{1}{c||}{Any} \\
 \hline\hline
Random Data 1
& 0.0	& 0.0	& 0.0	
& 0.0	& 0.0	& 0.0	
& 0.0	& 0.0	& 0.0	
& 0.0 \\
\hline
Random Data 2
& 0.0	& 0.0	& 0.0	
& 0.0	& 0.0	& 0.0	
& 0.0	& 0.0	& 0.0	
& 0.0 \\
\hline
Random Data 3
& 0.0	& 0.0	& 0.0	
& 0.0	& 0.0	& 0.0	
& 0.0	& 0.0	& 0.0	
& 0.0 \\
\hline
Random Data 4
& 0.0	& 0.0	& 0.0	
& 0.0	& 0.0	& 0.0	
& 0.0	& 0.0	& 0.0	
& 0.0 \\
\hline
Random Data 5
& 0.0	& 0.0	& 0.0	
& 0.0	& 0.0	& 0.0	
& 0.0	& 0.0	& 0.0	
& 0.0 \\
\hline
Random Data 6
& 0.0	& 0.0	& 96.1	
& 0.0	& 0.0	& 0.0	
& 0.0	& 0.0	& 0.0	
& 0.0 \\
\hline
Random Data 7
& 75.3	& 0.0	& 0.0	
& 0.0	& 0.0	& 0.0	
& 0.0	& 0.0	& 0.0	
& 0.0 \\
\hline
Random Data 8
& 0.0	& 0.0	& 0.0	
& 0.0	& 0.0	& 0.0	
& 0.0	& 0.0	& 0.0	
& 0.0 \\
\hline
Random Data 9
& 0.0	& 0.0	& 0.0	
& 0.0	& 0.0	& 0.0	
& 0.0	& 0.0	& 0.0	
& 0.0 \\
\hline
Random Data 10
& 0.0	& 0.0	& 0.0	
& 0.0	& 0.0	& 0.0	
& 0.0	& 0.0	& 0.0	
& 0.0 \\
\hline
\end{tabular}

\newpage
FitzHugh-Nagumo
\vskip 0.5em
 \begin{tabular}{||l||r|r|r|r|r|r|r|r|r|c||}
 \hline
 \multirow{2}{*}{Software}&
 \multicolumn{3}{c|}{\multirow{2}{*}{IQM}} &
 \multicolumn{3}{c|}{\multirow{2}{*}{SciML}} &
 \multicolumn{3}{c|}{\multirow{2}{*}{AMIGO2}} &
 Parameter\\
 & 
 \multicolumn{1}{c}{}& \multicolumn{1}{c}{} &\multicolumn{1}{c|}{}&
 \multicolumn{1}{c}{}& \multicolumn{1}{c}{} &\multicolumn{1}{c|}{}&
 \multicolumn{1}{c}{}& \multicolumn{1}{c}{} &\multicolumn{1}{c|}{}&
 Estimation.jl\\
 \hline
 Search Range  &
 \multicolumn{1}{c|}{[0,1]} & \multicolumn{1}{c|}{[0,2]} & \multicolumn{1}{c|}{[0,3]} &
 \multicolumn{1}{c|}{[0,1]} & \multicolumn{1}{c|}{[0,2]} & \multicolumn{1}{c|}{[0,3]} &
 \multicolumn{1}{c|}{[0,1]} & \multicolumn{1}{c|}{[0,2]} & \multicolumn{1}{c|}{[0,3]} &
 \multicolumn{1}{c||}{Any} \\
 \hline\hline
Random Data 1
& 82.0	& 165.1	& 128.0	
& 9.3	& 6.5	& 15.4	
& 0.0	& 0.0	& 0.0	
& 0.0 \\
\hline
Random Data 2
& 71.0	& 164.9	& 293.3	
& 0.2	& 0.0	& 0.0	
& 0.0	& 0.0	& 0.0	
& 0.0 \\
\hline
Random Data 3
& 58.0	& 58.0	& 58.0	
& 0.3	& 1.1	& 0.1	
& 0.0	& 0.0	& 0.0	
& 0.0 \\
\hline
Random Data 4
& 228.3	& 243.5	& 237.2	
& 0.0	& 0.0	& 0.0	
& 0.0	& 0.0	& 0.0	
& 0.0 \\
\hline
Random Data 5
& 66.6	& 105.7	& 105.7	
& 0.9	& 0.3	& 0.2	
& 0.0	& 0.0	& 0.0	
& 0.0 \\
\hline
Random Data 6
& 55.9	& 174.2	& 274.2	
& 0.0	& 0.0	& 0.0	
& 0.0	& 0.0	& 0.0	
& 0.0 \\
\hline
Random Data 7
& 81.2	& 81.2	& 88.0	
& 70.2	& 143.7	& 178.1	
& 0.0	& 0.0	& 2.2	
& 0.0 \\
\hline
Random Data 8
& 35.8	& 113.6	& 118.8	
& 0.9	& 0.1	& 0.4	
& 0.0	& 0.0	& 0.0	
& 0.0 \\
\hline
Random Data 9
& 101.3	& 191.5	& 191.5	
& 1.3	& 7.3	& 9.0	
& 0.3	& 0.3	& 0.3	
& 0.0 \\
\hline
Random Data 10
& 43.8	& 13.2	& 608.2	
& 0.0	& 0.0	& 0.3	
& 0.0	& 0.0	& 0.0	
& 0.0 \\
\hline
\end{tabular}

\vskip 2em

\centering
HIV
\vskip 0.5em
 \begin{tabular}{||l||r|r|r|r|r|r|r|r|r|c||}
 \hline
 \multirow{2}{*}{Software}&
 \multicolumn{3}{c|}{\multirow{2}{*}{IQM}} &
 \multicolumn{3}{c|}{\multirow{2}{*}{SciML}} &
 \multicolumn{3}{c|}{\multirow{2}{*}{AMIGO2}} &
 Parameter\\
 & 
 \multicolumn{1}{c}{}& \multicolumn{1}{c}{} &\multicolumn{1}{c|}{}&
 \multicolumn{1}{c}{}& \multicolumn{1}{c}{} &\multicolumn{1}{c|}{}&
 \multicolumn{1}{c}{}& \multicolumn{1}{c}{} &\multicolumn{1}{c|}{}&
 Estimation.jl\\
 \hline
 Search Range  &
 \multicolumn{1}{c|}{[0,1]} & \multicolumn{1}{c|}{[0,2]} & \multicolumn{1}{c|}{[0,3]} &
 \multicolumn{1}{c|}{[0,1]} & \multicolumn{1}{c|}{[0,2]} & \multicolumn{1}{c|}{[0,3]} &
 \multicolumn{1}{c|}{[0,1]} & \multicolumn{1}{c|}{[0,2]} & \multicolumn{1}{c|}{[0,3]} &
 \multicolumn{1}{c||}{Any} \\
 \hline\hline
Random Data 1
& 74.5	& 90.0	& 67.6	
& 2.9	& 7.8	& 11.9	
& 0.0	& 0.0	& 0.0	
& 0.0 \\
\hline
Random Data 2
& 125.8	& 113.5	& 119.2	
& 0.8	& 0.1	& 310.2	
& 0.0	& 0.0	& 0.0	
& 0.0 \\
\hline
Random Data 3
& 74.7	& 89.3	& 148.4	
& 45.2	& 8.1	& 18.8	
& 0.0	& 0.0	& 0.0	
& 0.0 \\
\hline
Random Data 4
& 66.4	& 185.7	& 264.6	
& 0.5	& 0.0	& 0.0	
& 0.0	& 0.0	& 0.0	
& 0.0 \\
\hline
Random Data 5
& 74.3	& 81.9	& 91.7	
& 0.1	& 151.7	& 151.8	
& 0.0	& 0.0	& 143.1	
& 0.0 \\
\hline
Random Data 6
& 80.1	& 68.2	& 91.5	
& 44.2	& 79.8	& 96.8	
& 184.3	& 4.6	& 0.0	
& 0.0 \\
\hline
Random Data 7
& 76.4	& 86.9	& 214.2	
& 40.2	& 43.1	& 306.5	
& 0.5	& 297.7	& 0.0	
& 0.0 \\
\hline
Random Data 8
& 89.1	& 118.7	& 204.4	
& 1.0	& 0.5	& 0.4	
& 0.0	& 0.0	& 0.0	
& 0.0 \\
\hline
Random Data 9
& 67.6	& 82.3	& 80.0	
& 12.4	& 78.3	& 131.8	
& 0.0	& 130.2	& 140.5	
& 0.0 \\
\hline
Random Data 10
& 127.4	& 125.8	& 165.3	
& 0.5	& 0.1	& 0.7	
& 0.0	& 0.0	& 0.0	
& 0.0 \\
\hline
\end{tabular}
\vskip 2em

\centering
Mammillary 3
\vskip 0.5em
 \begin{tabular}{||l||r|r|r|r|r|r|r|r|r|c||}
 \hline
 \multirow{2}{*}{Software}&
 \multicolumn{3}{c|}{\multirow{2}{*}{IQM}} &
 \multicolumn{3}{c|}{\multirow{2}{*}{SciML}} &
 \multicolumn{3}{c|}{\multirow{2}{*}{AMIGO2}} &
 Parameter\\
 & 
 \multicolumn{1}{c}{}& \multicolumn{1}{c}{} &\multicolumn{1}{c|}{}&
 \multicolumn{1}{c}{}& \multicolumn{1}{c}{} &\multicolumn{1}{c|}{}&
 \multicolumn{1}{c}{}& \multicolumn{1}{c}{} &\multicolumn{1}{c|}{}&
 Estimation.jl\\
 \hline
 Search Range  &
 \multicolumn{1}{c|}{[0,1]} & \multicolumn{1}{c|}{[0,2]} & \multicolumn{1}{c|}{[0,3]} &
 \multicolumn{1}{c|}{[0,1]} & \multicolumn{1}{c|}{[0,2]} & \multicolumn{1}{c|}{[0,3]} &
 \multicolumn{1}{c|}{[0,1]} & \multicolumn{1}{c|}{[0,2]} & \multicolumn{1}{c|}{[0,3]} &
 \multicolumn{1}{c||}{Any} \\
 \hline\hline
Random Data 1
& 67.0	& 67.2	& 80.3	
& 0.5	& 12.5	& 9.4	
& 0.0	& 0.0	& 0.0	
& 0.0 \\
\hline
Random Data 2
& 78.2	& 85.4	& 98.1	
& 44.4	& 89.4	& 105.4	
& 0.0	& 0.0	& 0.0	
& 0.0 \\
\hline
Random Data 3
& 89.1	& 104.1	& 69.5	
& 6.0	& 4.3	& 12.6	
& 0.0	& 0.0	& 0.0	
& 0.0 \\
\hline
Random Data 4
& 51.6	& 51.6	& 51.6	
& 7.2	& 13.2	& 29.1	
& 0.0	& 0.0	& 0.0	
& 0.0 \\
\hline
Random Data 5
& 57.9	& 60.1	& 102.9	
& 1.1	& 20.0	& 39.2	
& 0.0	& 0.0	& 0.0	
& 0.0 \\
\hline
Random Data 6
& 78.8	& 98.2	& 374.3	
& 0.4	& 0.6	& 4.6	
& 0.0	& 0.0	& 0.0	
& 0.0 \\
\hline
Random Data 7
& 111.6	& 300.4	& 71.8	
& 4.4	& 6.3	& 5.3	
& 0.0	& 0.0	& 0.0	
& 0.0 \\
\hline
Random Data 8
& 93.4	& 136.7	& 275.4	
& 5.1	& 17.1	& 10.5	
& 0.0	& 0.0	& 0.0	
& 0.0 \\
\hline
Random Data 9
& 70.4	& 90.8	& 90.8	
& 16.9	& 3.7	& 11.0	
& 0.0	& 0.0	& 0.0	
& 0.0 \\
\hline
Random Data 10
& 63.3	& 48.7	& 67.1	
& 33.4	& 47.5	& 20.2	
& 0.0	& 0.0	& 0.0	
& 0.0 \\
\hline
\end{tabular}
}

\newpage
{
Lotka-Volterra
\vskip 0.5em
 \begin{tabular}{||l||r|r|r|r|r|r|r|r|r|c||}
 \hline
 \multirow{2}{*}{Software}&
 \multicolumn{3}{c|}{\multirow{2}{*}{IQM}} &
 \multicolumn{3}{c|}{\multirow{2}{*}{SciML}} &
 \multicolumn{3}{c|}{\multirow{2}{*}{AMIGO2}} &
 Parameter\\
 & 
 \multicolumn{1}{c}{}& \multicolumn{1}{c}{} &\multicolumn{1}{c|}{}&
 \multicolumn{1}{c}{}& \multicolumn{1}{c}{} &\multicolumn{1}{c|}{}&
 \multicolumn{1}{c}{}& \multicolumn{1}{c}{} &\multicolumn{1}{c|}{}&
 Estimation.jl\\
 \hline
 Search Range  &
 \multicolumn{1}{c|}{[0,1]} & \multicolumn{1}{c|}{[0,2]} & \multicolumn{1}{c|}{[0,3]} &
 \multicolumn{1}{c|}{[0,1]} & \multicolumn{1}{c|}{[0,2]} & \multicolumn{1}{c|}{[0,3]} &
 \multicolumn{1}{c|}{[0,1]} & \multicolumn{1}{c|}{[0,2]} & \multicolumn{1}{c|}{[0,3]} &
 \multicolumn{1}{c||}{Any} \\
 \hline\hline
Random Data 1
& 80.1	& 80.1	& 80.1	
& 11.5	& 29.6	& 55.7	
& 0.0	& 166.6	& 0.0	
& 0.0 \\
\hline
Random Data 2
& 64.2	& 77.4	& 77.4	
& 19.7	& 28.0	& 42.9	
& 0.7	& 0.0	& 0.0	
& 0.0 \\
\hline
Random Data 3
& 64.1	& 75.6	& 75.6	
& 45.3	& 72.2	& 95.4	
& 0.0	& 59.0	& 5.3	
& 0.0 \\
\hline
Random Data 4
& 76.1	& 76.1	& 76.1	
& 262.8	& 242.1	& 229.5	
& 0.0	& 0.0	& 3.8	
& 0.0 \\
\hline
Random Data 5
& 70.7	& 65.8	& 65.8	
& 22.1	& 68.1	& 112.2	
& 0.0	& 0.0	& 0.0	
& 0.0 \\
\hline
Random Data 6
& 76.8	& 76.8	& 76.8	
& 43.1	& 41.9	& 44.6	
& 0.0	& 0.0	& 0.0	
& 0.0 \\
\hline
Random Data 7
& 77.1	& 77.1	& 77.1	
& 106.0	& 187.2	& 296.0	
& 66.9	& 749.2	& 0.0	
& 0.0 \\
\hline
Random Data 8
& 69.4	& 80.9	& 80.9	
& 4.5	& 8.7	& 23.7	
& 64.4	& 150.8	& 249.4	
& 0.0 \\
\hline
Random Data 9
& 70.4	& 71.1	& 71.1	
& 10.2	& 89.6	& 192.5	
& 0.0	& 0.0	& 2.1	
& 0.0 \\
\hline
Random Data 10
& 72.7	& 72.7	& 72.7	
& 22.0	& 1.6	& 24.3	
& 4.2	& 0.0	& 36.9	
& 0.0 \\
\hline
\end{tabular}

\vskip 2em

Crauste
\vskip 0.5em
 \begin{tabular}{||l||r|r|r|r|r|r|r|r|r|c||}
 \hline
 \multirow{2}{*}{Software}&
 \multicolumn{3}{c|}{\multirow{2}{*}{IQM}} &
 \multicolumn{3}{c|}{\multirow{2}{*}{SciML}} &
 \multicolumn{3}{c|}{\multirow{2}{*}{AMIGO2}} &
 Parameter\\
 & 
 \multicolumn{1}{c}{}& \multicolumn{1}{c}{} &\multicolumn{1}{c|}{}&
 \multicolumn{1}{c}{}& \multicolumn{1}{c}{} &\multicolumn{1}{c|}{}&
 \multicolumn{1}{c}{}& \multicolumn{1}{c}{} &\multicolumn{1}{c|}{}&
 Estimation.jl\\
 \hline
 Search Range  &
 \multicolumn{1}{c|}{[0,1]} & \multicolumn{1}{c|}{[0,2]} & \multicolumn{1}{c|}{[0,3]} &
 \multicolumn{1}{c|}{[0,1]} & \multicolumn{1}{c|}{[0,2]} & \multicolumn{1}{c|}{[0,3]} &
 \multicolumn{1}{c|}{[0,1]} & \multicolumn{1}{c|}{[0,2]} & \multicolumn{1}{c|}{[0,3]} &
 \multicolumn{1}{c||}{Any} \\
 \hline\hline
Random Data 1
& 86.9	& 119.2	& 150.4	
& 12.2	& 27.0	& 39.7	
& 0.0	& 0.0	& 81.1	
& 0.0 \\
\hline
Random Data 2
& 171.1	& 239.7	& 238.1	
& 22.1	& 82.3	& 135.8	
& 0.0	& 53.9	& 132.3	
& 0.0 \\
\hline
Random Data 3
& 78.7	& 66.2	& 467.4	
& 61.5	& 121.2	& 203.0	
& 13.8	& 113.2	& 0.0	
& 0.0 \\
\hline
Random Data 4
& 134.6	& 109.6	& 129.2	
& 30.8	& 69.5	& 130.2	
& 12.0	& 0.0	& 59.1	
& 0.0 \\
\hline
Random Data 5
& 92.7	& 189.2	& 66.2	
& 63.1	& 154.3	& 261.7	
& 0.0	& 101.2	& 0.0	
& 0.0 \\
\hline
Random Data 6
& 80.6	& 167.0	& 84.7	
& 16.4	& 35.2	& 484.7	
& 0.0	& 0.0	& 49.1	
& 0.0 \\
\hline
Random Data 7
& 83.6	& 78.0	& 83.0	
& 9.4	& 32.8	& 69.7	
& 0.0	& 0.0	& 0.0	
& 0.0 \\
\hline
Random Data 8
& 82.5	& 108.9	& 86.6	
& 11.9	& 65.0	& 117.0	
& 0.0	& 127.8	& 195.0	
& 0.0 \\
\hline
Random Data 9
& 125.2	& 101.6	& 304.6	
& 42.8	& 120.7	& 153.9	
& 0.0	& 248.6	& 0.0	
& 0.0 \\
\hline
Random Data 10
& 59.8	& 72.6	& 187.2	
& 203.0	& 21.1	& 668.3	
& 0.0	& 412.3	& 0.0	
& 0.0 \\
\hline
\end{tabular}
\vskip 2em

Biohydrogenation
\vskip 0.5em
 \begin{tabular}{||l||r|r|r|r|r|r|r|r|r|c||}
 \hline
 \multirow{2}{*}{Software}&
 \multicolumn{3}{c|}{\multirow{2}{*}{IQM}} &
 \multicolumn{3}{c|}{\multirow{2}{*}{SciML}} &
 \multicolumn{3}{c|}{\multirow{2}{*}{AMIGO2}} &
 Parameter\\
 & 
 \multicolumn{1}{c}{}& \multicolumn{1}{c}{} &\multicolumn{1}{c|}{}&
 \multicolumn{1}{c}{}& \multicolumn{1}{c}{} &\multicolumn{1}{c|}{}&
 \multicolumn{1}{c}{}& \multicolumn{1}{c}{} &\multicolumn{1}{c|}{}&
 Estimation.jl\\
 \hline
 Search Range  &
 \multicolumn{1}{c|}{[0,1]} & \multicolumn{1}{c|}{[0,2]} & \multicolumn{1}{c|}{[0,3]} &
 \multicolumn{1}{c|}{[0,1]} & \multicolumn{1}{c|}{[0,2]} & \multicolumn{1}{c|}{[0,3]} &
 \multicolumn{1}{c|}{[0,1]} & \multicolumn{1}{c|}{[0,2]} & \multicolumn{1}{c|}{[0,3]} &
 \multicolumn{1}{c||}{Any} \\
 \hline\hline
Random Data 1
& 83.6	& 120.3	& 132.8	
& 16.9	& 31.5	& 206.3	
& 0.0	& 0.0	& 0.0	
& 0.0 \\
\hline
Random Data 2
& 76.7	& 372.9	& 69.7	
& 69.1	& 161.5	& 646.2	
& 0.0	& 34.2	& 44.4	
& 0.0 \\
\hline
Random Data 3
& 72.5	& 110.5	& 72.6	
& 169.6	& 24.8	& 67.5	
& 0.0	& 0.0	& 0.0	
& 0.0 \\
\hline
Random Data 4
& 66.8	& 90.7	& 169.5	
& 38.7	& 38.1	& 341.5	
& 0.0	& 0.0	& 0.0	
& 0.1 \\
\hline
Random Data 5
& 80.9	& 411.8	& 213.4	
& 33.7	& 303.8	& 181.6	
& 43.0	& 0.0	& 66.9	
& 0.0 \\
\hline
Random Data 6
& 107.6	& 85.1	& 73.4	
& 22.7	& 59.7	& 359.8	
& 0.0	& 53.4	& 0.0	
& 0.0 \\
\hline
Random Data 7
& 171.5	& 381.4	& 344.3	
& 132.4	& 407.4	& 607.8	
& 0.0	& 0.0	& 0.0	
& 0.0 \\
\hline
Random Data 8
& 81.5	& 63.1	& 69.0	
& 147.5	& 214.3	& 227.7	
& 62.3	& 0.0	& 48.8	
& 0.0 \\
\hline
Random Data 9
& 118.4	& 116.4	& 80.2	
& 111.0	& 223.9	& 395.3	
& 72.9	& 47.8	& 48.1	
& 0.0 \\
\hline
Random Data 10
& 78.0	& 88.4	& 77.6	
& 35.7	& 45.7	& 31.8	
& 0.0	& 0.0	& 75.1	
& 0.0 \\
\hline
\end{tabular}


\newpage
Mammillary 4
\vskip 0.5em
 \begin{tabular}{||l||r|r|r|r|r|r|r|r|r|c||}
 \hline
 \multirow{2}{*}{Software}&
 \multicolumn{3}{c|}{\multirow{2}{*}{IQM}} &
 \multicolumn{3}{c|}{\multirow{2}{*}{SciML}} &
 \multicolumn{3}{c|}{\multirow{2}{*}{AMIGO2}} &
 Parameter\\
 & 
 \multicolumn{1}{c}{}& \multicolumn{1}{c}{} &\multicolumn{1}{c|}{}&
 \multicolumn{1}{c}{}& \multicolumn{1}{c}{} &\multicolumn{1}{c|}{}&
 \multicolumn{1}{c}{}& \multicolumn{1}{c}{} &\multicolumn{1}{c|}{}&
 Estimation.jl\\
 \hline
 Search Range  &
 \multicolumn{1}{c|}{[0,1]} & \multicolumn{1}{c|}{[0,2]} & \multicolumn{1}{c|}{[0,3]} &
 \multicolumn{1}{c|}{[0,1]} & \multicolumn{1}{c|}{[0,2]} & \multicolumn{1}{c|}{[0,3]} &
 \multicolumn{1}{c|}{[0,1]} & \multicolumn{1}{c|}{[0,2]} & \multicolumn{1}{c|}{[0,3]} &
 \multicolumn{1}{c||}{Any} \\
 \hline\hline
Random Data 1
& 54.0	& 73.2	& 111.7	
& 15.7	& 26.0	& 116.4	
& 32.8	& 41.2	& 29.7	
& 0.4 \\
\hline
Random Data 2
& 63.4	& 235.1	& 87.6	
& 213.9	& 2.5	& 2.4	
& 0.0	& 0.0	& 0.0	
& 0.0 \\
\hline
Random Data 3
& 72.0	& 58.1	& 65.8	
& 25.0	& 5.3	& 6.2	
& 0.0	& 74.1	& 74.1	
& 0.0 \\
\hline
Random Data 4
& 62.5	& 147.1	& 55.5	
& 53.6	& 53.6	& 68.1	
& 137.7	& 106.3	& 144.1	
& 0.2 \\
\hline
Random Data 5
& 87.8	& 89.9	& 156.5	
& 12.9	& 129.0	& 191.7	
& 0.0	& 26.1	& 26.1	
& 0.0 \\
\hline
Random Data 6
& 79.2	& 86.8	& 172.7	
& 16.7	& 157.7	& 190.0	
& 9.8	& 66.3	& 0.0	
& 0.0 \\
\hline
Random Data 7
& 108.3	& 98.4	& 130.2	
& 40.2	& 40.2	& 10.9	
& 118.1	& 0.0	& 109.4	
& 0.0 \\
\hline
Random Data 8
& 183.1	& 159.8	& 108.4	
& 66.9	& 68.6	& 408.2	
& 0.0	& 144.7	& 123.8	
& 0.0 \\
\hline
Random Data 9
& 157.8	& 84.5	& 63.3	
& 124.7	& 95.3	& 95.3	
& 0.0	& 89.0	& 0.0	
& 0.0 \\
\hline
Random Data 10
& 72.9	& 63.0	& 101.8	
& 92.8	& 99.0	& 99.0	
& 0.0	& 18.9	& 82.8	
& 0.0 \\
\hline
\end{tabular}
\vskip 2em

SEIR
\vskip 0.5em
 \begin{tabular}{||l||r|r|r|r|r|r|r|r|r|c||}
 \hline
 \multirow{2}{*}{Software}&
 \multicolumn{3}{c|}{\multirow{2}{*}{IQM}} &
 \multicolumn{3}{c|}{\multirow{2}{*}{SciML}} &
 \multicolumn{3}{c|}{\multirow{2}{*}{AMIGO2}} &
 Parameter\\
 & 
 \multicolumn{1}{c}{}& \multicolumn{1}{c}{} &\multicolumn{1}{c|}{}&
 \multicolumn{1}{c}{}& \multicolumn{1}{c}{} &\multicolumn{1}{c|}{}&
 \multicolumn{1}{c}{}& \multicolumn{1}{c}{} &\multicolumn{1}{c|}{}&
 Estimation.jl\\
 \hline
 Search Range  &
 \multicolumn{1}{c|}{[0,1]} & \multicolumn{1}{c|}{[0,2]} & \multicolumn{1}{c|}{[0,3]} &
 \multicolumn{1}{c|}{[0,1]} & \multicolumn{1}{c|}{[0,2]} & \multicolumn{1}{c|}{[0,3]} &
 \multicolumn{1}{c|}{[0,1]} & \multicolumn{1}{c|}{[0,2]} & \multicolumn{1}{c|}{[0,3]} &
 \multicolumn{1}{c||}{Any} \\
 \hline\hline
Random Data 1
& 99.7	& 140.4	& 204.1	
& 4.9	& 55.4	& 109.1	
& 0.0	& 12.1	& 0.0	
& 0.0 \\
\hline
Random Data 2
& 103.8	& 192.9	& 294.3	
& 24.7	& 26.0	& 62.1	
& 0.0	& 53.3	& 55.6	
& 0.0 \\
\hline
Random Data 3
& 296.0	& 720.9	& 1096.5	
& 24.2	& 20.0	& 59.0	
& 28.3	& 10.9	& 38.7	
& 0.0 \\
\hline
Random Data 4
& 73.8	& 112.7	& 161.6	
& 19.4	& 24.6	& 63.9	
& 33.6	& 1.7	& 65.4	
& 0.0 \\
\hline
Random Data 5
& 134.1	& 255.1	& 390.5	
& 10.1	& 43.7	& 101.8	
& 46.3	& 46.3	& 0.0	
& 0.0 \\
\hline
Random Data 6
& 82.8	& 133.1	& 198.1	
& 16.5	& 19.6	& 57.6	
& 7.4	& 24.1	& 22.9	
& 0.0 \\
\hline
Random Data 7
& 84.0	& 125.0	& 185.6	
& 45.3	& 122.7	& 178.6	
& 29.0	& 15.5	& 16.9	
& 0.0 \\
\hline
Random Data 8
& 129.5	& 180.4	& 266.3	
& 11.6	& 63.1	& 143.6	
& 0.0	& 75.8	& 74.8	
& 0.0 \\
\hline
Random Data 9
& 90.8	& 165.5	& 249.9	
& 17.7	& 100.1	& 165.2	
& 113.9	& 167.2	& 226.8	
& 0.0 \\
\hline
Random Data 10
& 228.4	& 281.6	& 342.3	
& 9.8	& 130.8	& 240.0	
& 0.0	& 0.0	& 7.1	
& 0.0 \\
\hline
\end{tabular}
}

\end{document}